\documentclass[mathleft
]{an}
\usepackage{graphicx}
\usepackage{times}
\begin{document}

\Pagespan{1}{}
\Yearpublication{2009}%
\Yearsubmission{2009}%
\Month{x}%
\Volume{x}%
\Issue{x}%

\hyphenation{pro-gress}

\title{X-ray observations of classical novae. Theoretical implications}

\author{M. Hernanz\inst{1}\fnmsep\thanks{Corresponding author:
  \email{hernanz@ieec.uab.es}\newline}
\and  G. Sala\inst{2}
}
\titlerunning{X-ray observations of classical novae. Theoretical implications}
\authorrunning{M. Hernanz and G. Sala}
\institute{Institut de Ci\`encies de l'Espai (CSIC-IEEC),
Campus UAB, Facultat de Ci\`encies, C5 parell 2$^{on}$,
08193 Bellaterra (Barcelona), Spain
\and 
Departament F\'{\i}sica i Enginyeria Nuclear, EUETIB (UPC-IEEC), 
Comte d'Urgell 187, 08036 Barcelona, Spain}

\received{}
\accepted{}
\publonline{}


\abstract{Detection of X-rays from classical novae, both in outburst and post-outburst, provides unique 
and crucial information 
about the explosion mechanism. Soft X-rays reveal the hot white dwarf photosphere, whenever hydrogen (H)
nuclear burning is still on and expanding envelope is transparent enough, 
whereas harder X-rays give information about the ejecta and/or the accretion 
flow in the reborn 
cataclysmic variable. The duration of the supersoft X-ray emission phase is related to the turn-off of the 
classical nova, i.e., of 
the H-burning on top of the white dwarf core. A review of X-ray observations is presented, with a special 
emphasis on the 
implications for the duration of post-outburst steady H-burning and its theoretical explanation. The particular 
case of recurrent 
novae (both the "standard" objects and the recently discovered ones) will also be reviewed, in terms of 
theoretical feasibility of 
short recurrence periods, as well as regarding implications for scenarios of type Ia supernovae.
}

\maketitle

\section{Introduction}
White dwarfs in cataclysmic variables explode as classical novae, when they accrete H-rich matter from their 
main sequence companion with some particular combinations of mass-accretion rate and initial white dwarf 
mass and luminosity. Matter accumulates on top of the white dwarf until hydrogen ignites, 
under degenerate or semi-degenerate conditions.This leads to a thermonuclear runaway, 
because self adjustment of the envelope once nuclear burning starts is not possible. 

As a consequence of the explosion, a fraction of the accreted envelope (or even the whole envelope plus some material 
dredged-up from the white dwarf core) is ejected, with mass typically in the range $\sim10^{-6}-10^{-4}$ M$_\odot$, 
velocity of 100's or 1000's of km/s, and luminosity up to around $\sim 10^{4-5}$ L$_\odot$ (Prialnik \& Kovetz 1995, 
Starrfield et al. 1998, Jos\'e \& Hernanz 1998). Steady nuclear burning is expected to take place in the 
remaining (if existing) H-rich envelope on top of the white dwarf.
In fact, multiwavelength observations of classical novae have shown that a 
phase of constant bolometric luminosity (with values close to the Eddington limit) occurs after 
maximum, during the early optical decline; as the envelope mass is depleted, the photospheric radius decreases and 
thus the effective temperature increases, leading to a hardening of the spectrum, from optical 
to ultraviolet, extreme ultraviolet and soft X-rays, in agreement with theoretical predictions (Starrfield 1989, 
MacDonald 1996, Krautter 2002). Therefore, novae are expected to emit soft X-rays (E$\le$1 keV) when steady 
H-bur\-ning at the base of the white dwarf (WD) envelope heats the WD photosphere to effective temperatures from
a few $10^5$ K up to $10^6$ K (Kato \& Hachisu 1994, Sala \& Hernanz 2005a, Kato these 
proceedings), provided that the expanding envelope is not opaque to X-rays; 
the nova then behaves as a supersoft source, SSS. However, if no 
H-rich envelope is left, or if the steady H-burning phase ends-up before the ejecta becomes optically thin
to soft X-rays, 
the SSS phase will be absent. But such very short duration steady H-burning phases correspond to very tiny 
H-rich envelopes left after the explosion, much smaller than predicted by hydrodynamical models of nova 
explosions.
 
From the observation of the soft X-ray emission, the 
duration of the turn-off of the nova explosion can be measured; some other properties, such as 
the mass and metallicitof the remnant H-burning envelope and the mass of the underlying WD can be constrained, 
by comparing observations with current post-outburst nova models of this  phase (Sala \& Hernanz 2005a). 
Additional information about the decline of classical 
novae comes from UV observations (Shore et al. 1996, G\'onzalez-Riestra et al. 
1998): turn-off times derived from UV data were in agreement with the soft X-ray ones, when both data 
were available.

There are other mechanisms of X-ray emission in classical 
novae, not related to residual H-nuclear burning on top of the white dwarf. 
Internal 
shocks in the expanding envelope, as well as shocks between the ejecta and circumstellar matter, lead 
to the heating of the plasma and the ensuing emission of X-rays, mainly by thermal bremsstrahlung; the 
corresponding energy of the emitted photons is larger than the energy of supersoft X-rays from the hot white 
dwarf photosphere. Such hard X-ray emission has been detected early after the explosion 
in some novae, e.g., V838 Her 1991 (5 days after optical outburst, which is the earliest 
detection of X-rays from a classical nova during outburst, Lloyd et al. 1992). In other novae it was 
detected later, e.g., 16 months after outburst in V 351 Pup 1991 (Orio et al. 1996), and both early and late 
in V382 Vel 1999 (15 days and 6 months after outburst, Orio et al. 2001). In the particular case of 
the recurrent nova RS Oph, hard X-rays up to $\sim 50$ keV were detected very early after its 2006
outburst (Sokoloski et al. 2006, Bode et al. 2006), 
revealing the interaction between the ejecta and the wind of the red giant companion (RS Oph 
did not explode in a cataclysmic variable, as classical novae do; see next section). 

It has also been suggested that the Comptonisation of the gamma-rays produced in $^{22}$Na nuclear 
decay could be responsible of prompt hard X-rays emitted by novae (Livio et al. 1992); but this 
has been ruled out after detailed modelling of the gamma-ray spectra 
of classical novae: Compton down-scattered photons are harder 
(energies larger than 20-30 keV, because of photoelectric absorption) and 
emitted earlier than the observed hard X-rays. In fact, emission of comptonised gamma-rays 
has very short duration (less than 2 days) and occurs before the nova is discovered optically 
(G\'omez-Gomar et al 1998, Hernanz et al. 1999). The recent eruption of the 
recurrent nova RS Oph in 2006 was a good example of an early hard X-ray emission 
(detected with Swift/BAT and RXTE, Bode et al. 2006, Sokoloski et al. 2006, see above) 
non attributable to such Comptonisation of gamma-rays (Hernanz \& Jos\'e 2008a).  

Finally, post-outburst novae also emit X-rays once accretion is reestablished, behaving then 
like cataclysmic variables (see for instance the case of V2487 Oph 1998, 
Hernanz \& Sala 2002).

Hard X-ray emission itself is out of the scope of this review, but it should be mentioned that in some 
cases it is difficult to disentangle which fraction of the soft part of a whole X-ray spectrum corresponds to 
thermal plasma emission and which one to hot white dwarf photospheric emission. Only when 
``canonical" supersoft emission  is clearly seen, with large luminosities close 
to the Eddington limit and reasonable effective temperatures (up to about 100 eV), can 
it be ascribed to the whole hot white dwarf photosphere. 

\section{Observational results}

The first nova detected in soft X-rays was GQ Mus 1983, 
observed with EXOSAT in 1983, 460 days after optical 
maximum (\"Ogelman et al. 1984). In the 90's, the 
R\"oentgen satellite, ROSAT, provided new and better quality data of novae, specially 
in the soft X-ray range, thanks to the PSPC instrument. But only 3 novae 
(V1974 Cyg 1992, GQ Mus 1983 and N LMC 1995) 
were found emitting soft X-rays, from a sample of 30 galactic and 9 LMC novae observed up 
to 10 years after their explosion, both during the ROSAT All Sky Survey 
and in pointed observations (Orio et al. 2001). GQ Mus 1983 was the nova 
with the longest supersoft X-ray phase, 9 years (\"Ogelman et al. 1993, 
Shanley et al. 1995, Orio et al. 2001, Balman \& Krautter 2001). V1974 Cyg 1992 
was a very bright nova, observed at all wavelengths; a complete soft X-rays 
light curve was obtained with ROSAT, from the early rise phase to the plateau 
and the decline, lasting 1.5 years (Krautter et al. 1996, Balman et al. 1998). 
The third and last nova discovered in soft X-rays with ROSAT was N LMC 
1995 (Orio \& Greiner 1999). 

After the ROSAT era, observations with Beppo\-SAX, active during the period 1996-2003, 
and with Chandra and XMM-Newton (both launched in 1999) have 
revealed soft X-ray emission in a few more classical novae, and in the recent eruption of the 
recurrent nova RS Oph in 2006. 
A wealth of emission lines and puzzling temporal behaviours have been deduced. 
Beppo\-SAX observations of V382 Vel 1999 showed a non constant (during 
the observation) supersoft X-ray flux, which could not be fit with a model atmosphere, 
because there were emission lines from highly ionised species superimposed on the continuum 
(Orio et al. 2002). Later Chandra grating observations showed that the source had faded 
considerably, shifting to a pure emission line spectrum; the turn-off of soft X-ray emission, and 
thus of hydrogen nuclear burning, occurred between 7 and 9 months after the explosion 
(Burwitz  et al. 2002, Ness et al. 2003). 

Two other interesting objects observed with Chandra we\-re V1494 (N Aql 1999 No.2)  
and V4743 Sgr (Nova Sgr 2002 No. 3). V1494 Aql showed a 
strong soft component 10 months after the explosion (which was not present 6 months after 
outburst), with a puzzling light curve, never seen before in a nova: a short timescale 
rise and decline (burst) and oscillations, that may be attributed to 
pulsations of the white dwarf (Drake et al 2003). V4743 Sgr  
was observed with Chandra 6.5 months after outburst, displaying extremely bright 
supersoft X-ray 
emission, with again a very variable light curve (Ness et al. 2003, Leibowitz et al. 2006). 
These two novae and V382 Vel were the first to show puzzling temporal variability 
during their supersoft X-ray emission phase -or a bit later; however, nowadays variability seems to be the 
rule and not the exception (e.g., in recent novae observed with Swift), indicating that the advent of better 
instruments has opened a new window to the study the nova phenomenon. Each new nova has its own 
characteristics, and does not behave like other previous observed ones. So a larger 
sample is required to define classes and understand the underlying physics.

A monitoring campaign of young post-outburst galactic classical novae
with XMM-Newton in 2000-2001, before Swift satellite launch in 2004, aimed to determine 
the duration of 
the turn-off phase and to increase the data base of novae emitting as SSS, then including 
very few novae (Hernanz et al. 2008). 
Two novae were clearly detected: V2487 Oph 1998 and V4633 Sgr 1998, 
but they were not SSSs and thus H-burning had already turned-off. V4633 Sgr 1998
showed an X-ray spectrum dominated by thermal plasma emission, from the shock-heated
expanding shell or from the reestablished accretion flow (Hernanz \& Sala 2007, Ferri et al. 2007, 
Hernanz et al. 2009). 
Observations of V2487 Oph 1998 provided clear evidence that accretion was reestablished 
as early as 2.7 years after the explosion (Hernanz \& Sala 2002). This conclusion 
stems mainly from the detection of a fluorescent Fe K$_\alpha$ emission line at 6.4 keV, 
indicating the existence of reflection on cold neutral matter, either the accretion disk 
(if any) or the WD surface. 

Later campaigns with XMM-Newton were more successful concerning the search for supersoft 
X-ray emission (Hernanz et al. 2008). 
Two novae were found in this phase: V5115 Sgr 2005a and V5116 Sgr 2005b, 
18 and 20 months after outburst. The EPIC spectra of V5115 Sgr 2005a are shown in 
Figure 1. An ONe white dwarf atmosphere with T$= 6\times 10^5$K and 
L$_{\rm WD}= 10^{36}$erg/s (for d=10 kpc) fits well the soft X-ray emission, but there is
additional emission at hard X-rays, corresponding to a hot plasma with kT$\sim 2.5$ keV, 
most probably revealing the ejecta. A later observation of 
this same nova in 2009 showed that there was no soft X-ray emission anymore, and thus 
H-burning had turned-off.
Hachisu \& Kato (2007) predicted a too short supersoft X-ray emitting phase for this nova, 
based on their optical thick wind model of light curves at several wavelengths; according 
to their model, it should have ended by day 250-280. 
Regarding V5116 Sgr 2005b, it was detected as a bright supersoft source 20 months after 
outburst (Sala et al. 2008). The X-ray light curve showed abrupt increases and decreases 
of the flux, by a factor $\sim 8$, with a tentative period consistent with the orbital 
period observed in the optical (Dobrotka et al. 2008). The emission was fitted with an 
ONe white dwarf atmosphere model, with an atmospheric temperature not changing 
between the low and high states. A new recent XMM-Newton observation of this nova in March 2009 
(44 months after outburst) has revealed that the source was not emitting supersoft X-rays anymore. 
The turn-off of H-burning thus took place between 2 and 3 years after outburst (see Sala et al. 
in these proceedings); Swift also observed this source off some months earlier (Osborne 2009). 
These turn-off times agree with those predicted by Hachisu \& Kato (2007). As a summary of the 
XMM-Newton campaigns, eleven novae were observed between 3 months and 5 years after outburst; 
four were not detected, two were detected marginally and four were clearly detected, 
but only two of them - V5115 Sgr 2005a and V5116 Sgr 2005b - were still emitting supersoft X-rays; 
the other three detected - V4533 Sgr 2008, V2487 Oph 2008 and V2362 Cyg 2006- revealed 
either reestablished accretion or hot shocked ejecta. 

The Swift satellite has contributed very significantly to the study of the X-ray 
emission from novae, through  
its campaign of novae monitoring with the XRT telescope and the UVOT monitor (Ness et al. 2007); it 
has also served as a trigger for ToO observations with XMM-Newton and Chandra 
(Osborne 2009 and these proceedings). There is a nova observed by Swift which 
deserves particular attention, regarding the length of the supersoft X-ray emission: V723 Cas 
(N Cas 1995). This nova was still bright in supersoft X-rays, when observed 
with Swift in 2006-2007, thus having the longest duration of supersoft X-ray emission phase 
(Ness et al. 2008). However, V723 Cas was quite dim and its Swift/XRT spectra had poor 
spectral resolution, so that the observations were not compelling and it was not 100\% clear that 
the origin of the emission was residual H-nuclear burning on top of the white dwarf; the fits gave too 
small luminosities and too large absorptions. Eventually, emission "cataclysmic variable-like" could 
be an alternative explanation: in these binaries, soft X-ray emission related to the 
heating of the white dwarf atmosphere as a consequence of accretion is often observed, 
specially if the system is magnetic (see the review by Cropper 1990). Several other novae have 
been detected by Swift as SSSs. 
According to Osborne (2009), 35 novae had been observed, earlier than 4000 days post outburst, 
with Swift (March 2009 data, see update in Osborne's contribution to these proceedings); 
19 were detected in X-rays, out of which 9 (including RS Oph 2006 outburst) were SSS. So 
ROSAT's (plus Beppo-SAX, XMM-Newton and Chandra) statistics have really improved thanks to Swift. 
The duration of the supersoft X-rays 
emitting phase, however, is still quite short: at most 3.5 years (V574 Pup 2004) for all 
novae detected as SSS except the above mentioned V723 Cas (N Cas 1995). 

\subsection{The recurrent nova RS Oph}

A really exceptional object is the recurrent nova RS Oph, which had its latest  
eruption in 2006, after its most recent previous one in 1985 (see Bode in these proceedings). 
The scenario of RS Oph is different 
from that of 
classical novae, because the companion is a red giant instead of a main sequence star, 
and the orbital period is much larger than the typical one for cataclysmic variables. 
Large mass accretion rates and very massive white dwarfs are required to power such frequent 
explosions, with a recurrence period of only 21 years, to be compared to the typical $10^{4-5}$ years of 
standard classical novae, as already pointed out long ago (Starrfield et al. 1985). 
Recent models of RS Oph indicate that the white dwarf mass and mass accretion rate should be 
larger than 1.35 M$_\odot$ and $10^{-8}$M$_\odot$/yr, respectively (Hernanz \& Jos\'e 2008b). 
An important result is that accreted masses are larger than ejected ones, so that the mass of the 
white dwarf is expected to increase in each eruption, making RS Oph a potential type Ia supernova 
candidate; however, such massive white dwarfs are expected to be of the ONe type, instead of CO, 
which poses a problem for the explosion scenario (Hernanz \& Jos\'e 2008b).

The 2006 eruption of RS Oph has been observed in depth in X-rays; there was both supersoft 
and hard (up to $\sim 50$ keV) X-ray emission. Early X-ray 
observations with RXTE (Sokoloski et al. 2006) and Swift/BAT (Bode et al. 2006) showed hard X-ray 
emission, revealing the blast wave from RS Oph, i.e., an outward 
propagating shock wave consequence of the interaction between the expanding ejecta and the 
red giant wind (much denser than typical circumstellar material in classical novae). 
The difference between the shock velocities deduced from 
X-ray and IR data indicated efficient particle acceleration (Tatischeff \& Hernanz 2007); 
nonlinear diffusive shock acceleration of cosmic rays in RS Oph also explained the 
deceleration of the blast wave in RS Oph, which occurred faster than predicted by the 
standard adiabatic shock wave model. But the most relevant 
data in the context of SSS came from Swift/XRT, XMM-Newton and Chandra observations; 
soft X-ray emission was detected at about 30 days post outburst, lasting only 60 days, 
with high temporal variability (Bode et al. 2006). The short duration of the supersoft X-ray 
phase pointed out to a very massive white dwarf in RS Oph (Hachisu et al. 2007), in agreement with 
the mass left after the explosion in recent theoretical models (Hernanz \& Jos\'e 2008b).
This has important implications, since RS Oph-type recurrent novae contain  
white dwarfs close to the Chandrasekhar mass limit, which increase their mass after each 
eruption; therefore, these are excellent type Ia supernovae candidates. 
Detailed analysis of the grating observations of RS Oph at different phases,including the supersoft 
X-ray one, show a wealth of emission features, often blueshifted, some of them still lacking 
theoretical interpretation (Ness et al. 2007, Nelson et al. 2008, Drake et al. 2009).

\subsection{M31}
Last but not least, the monitoring of our neighbour galaxy M31, with XMM-Newton and 
Chandra, has provided an incredibly large amount of new and exciting data (see Pietsch et al. 2007 
and review by Pietsch, and contribution by Henze in this workshop). Andromeda galaxy seems to be 
quite prolific in classical novae, often discovered trough their X-ray emission and later 
confirmed through detailed analysis of archival optical data.
 
\section{Theoretical implications}

The duration of the supersoft X-ray emitting phase is quite short in 
almost all detected novae: it is typically less than 1 year, with a few cases 
around 2-3.5 years, and three around 10 years 
(GQ Mus 1983: 9 years, N LMC 2005: about 8 years, and V723 Cas 1995:  
larger than 12 years). These values are, at first glance (but see below) 
hardly compatible with the evolution, on a purely nuclear time\-scale, of the  
H-rich material remaining on top of the white dwarf 
after explosive mass ejection. A look at some numbers
helps to understand this problem. In Table 1 the amount of mass 
left after the explosion and its chemical composition, for three
nova models computed by Jos\'e \& Hernanz (1998) for a 1.15 M$\odot$ 
initial WD mass, with a range of 
initial degrees of mixing between the WD core and accreted matter, are shown. 
$\Delta \rm M_{\rm rem}$ is the mass of the envelope left after the explosion:  
$\Delta \rm M_{\rm rem}=\Delta \rm M_{\rm env}-\Delta \rm M_{\rm ej}$ (see Yaron 
et al. 2005 for a complete set of CO nova models).  
Luminosity and $\tau_{\rm nuc}$ (time for turn-off of H-nuclear burning) 
come from the following expressions (from Sala \& Hernanz 2005a 
and Starrfield 1989, respectively):

$L^{plateau}_{ONe} (L_\odot) \simeq 5.95 \times 10^4 (\frac{M_c}{M_\odot}-0.536 X - 0.14)$

$\tau_{nuc} (yr) = 400 (\frac{M_H}{10^{-4}M_\odot})/(\frac{L}{2 \times 10^4 L_\odot})$

\noindent
where M$_{\rm H}$ is $\Delta M_{\rm rem} \times X$, with X the hydrogen mass fraction.
As seen in Table 1, the turn-off time ($\tau_{nuc}$) ranges from 
2 to 10 years, strongly depending on the chemical composition (as explained below). 
Around 1/3 to 1/2 of novae are expected to be of the ONe type (as the models in 
Table 1), according to 
abundance determinations of the ejecta (Livio \& Truran 1994). 
Then, why so few novae observed during the ROSAT and XMM-Newton 
monitoring campaigns described in previous section were detected in supersoft X-rays? 

An approach to the answer comes from a more specific study 
of steady H-burning on top of WDs (Sala and Hernanz 2005a). A grid of WD envelopes 
undergoing steady H-burning,   
with a range of realistic chemical compositions (taken from Jos\'e \& Hernanz 1998)
and of WD masses (0.9 to 1.3 M$_\odot$), 
was computed by Sala \& Hernanz (2005a). The loci 
of the models in the HR diagram shows that there are two branches, but just the 
high luminosity one corresponds to stable models (see their Figure 1). 
It is important to remark that the maximum effective temperatures reached depend solely 
on the WD mass and on the envelope chemical composition (mainly on its H-content X), 
ranging between $\sim 60$ and 100 eV; therefore, if 
the maximum T$_{\rm eff}$ attained by a nova can be measured through X-ray observations, 
a determination of M$_{\rm WD}$ plus the chemical composition of the remaining envelope 
could be done, independently of the luminosity, and thus of the distance. This measure 
is not straightforward, since the peak of T$_{\rm eff}$ is often missed. 
Another result is the determination of the ``plateau'' luminosity, which depends not only 
on the WD mass but also on the H-content of the envelope (see expression above and some 
values in Table 1). A relationship between envelope mass and effective temperature was 
also derived, which gives two 
important results: 1) the maximum T$_{\rm eff}$ reached are 100 eV, 
for very large WD masses, e.g., 
1.3 M$_\odot$; 2) envelope masses are very small, ranging from 
$ \sim 2 \times 10^{-7}$ to $ \sim 3 \times 10^{-6}$ M$_\odot$. 
This latest result was already obtained by Tuchman \& Truran (1998), who 
analysed the relevance of the chemical composition of the envelope left 
after nova explosions.
The envelope masses are much smaller 
than $\Delta \rm M_{\rm rem}$ shown in Table 1, and thus the time needed 
to get rid of them by steady burning, i.e. the turn-off time computed as 
$\tau_{nuc}$ above, is much shorter. An example: a ONe nova with 
M$_{\rm WD}=1.15$ M$_\odot$ and 50 \% initial mixing, reaches 
$L_{plateau}$ $=4.9 \times 10^4$ L$_\odot$, with a H-steady burning envelope 
of $7 \times 10^{-7}$ M$_\odot$, smaller than 
theoretically predicted by hydrodynamical models (see Table 1). The corresponding 
turn-off time $\tau_{nuc}$ is 0.4 years, instead of 7.4 years, in better agreement 
with observations.
Consequently, there is consistency with the predicted properties of WD envelopes 
steadily burning hydrogen in post-outburst novae (e.g., effective temperatures and 
luminosities) and the observations of supersoft X-ray emission. However, it is not 
yet known how the nova explosion can leave small enough  H-rich envelope masses; 
some extra me\-chanism of mass loss should be advocated to understand this difference. 

More details of the global properties of the steady burning H-rich envelopes are shown in 
Table 1 in Sala \& Hernanz (2005a). For instance, if 
quasi-static temporal evolution is assumed, the time needed for the envelope 
to change its mass by hydrogen burning only is the following:

$\Delta t = \epsilon \frac {\Delta M_{env} X_H}{L}$

\noindent
where $\epsilon$ is the nuclear energy released per gram of hydrogen burned 
($6 \times 10^{18}$ erg/g). A successful application of this model has been the 
analysis of the supersoft X-ray emission of V1974 Cyg 1992, which lasted about 
18 months (Krautter et al. 1996, Balman et al. 1998).  
Its evolution was modelled with a tiny  
envelope of only $\sim 2\times 10^{-6}$M$_\odot$, either on a 0.9 M$_{\odot}$ 
CO nova with 50 \% mixing, or in a 1.0 M$_{\odot}$ ONe nova with 25 \% mixing 
(Sala \& Hernanz 2005b), in good agreement with models of the optical 
and UV light curves (Hachisu \& Kato 2006). In fact, the CO nova option is in 
contradiction with the optical and UV spectral observations, which indicated 
that V1974 Cyg ejecta was enhanced in neon, and thus it was an ONe nova.

\section{Conclusions}

\begin{itemize}
\item
The number of novae detected as SSS has increased considerably in the last years, 
thanks mainly to Swift, XMM-Newton and Chandra. Some puzzling results concerning the 
temporal behaviour have been discovered, such as short period oscillations, 
sudden emergence and/or disappearance of soft X-ray emission.
More observations are still needed to 
establish a classification and understand such diversity.

\item 
Grating spectra are very rich, showing a wealth of lines; blueshifts of the absorption lines  
are specially relevant, since they reveal expansion during the super soft X-ray emitting 
phase (see Ness contribution to these proceedings). Detailed emission models 
(e.g., realistic WD atmospheres accounting for expansion effects) are in progress 
(see Van Rossum contribution to these proceedings).

\item
The WD mass and the envelope chemical composition - mainly its hydrogen content - determine the duration 
of the supersoft X-ray phase.

\item
Observed duration of the supersoft X-ray phase indicates, in general, 
M$_{\rm env}$ $<$ M$_{\rm accreted}$ - M$_{\rm ejected}$ from hydrodynamical models. 
The mechanism leading to such extra mass loss is not completely clear. 
Wind mass loss clearly contributes (see Kato \& Hachisu 1994 and subsequent papers).
However, up to now the evolutionary hydrodynamical models that follow the whole nova 
outburst up to the mass-loss phase, do not obtain 
the tiny M$_{\rm env}$ required, except for extreme combinations of 
accretion rate and initial white dwarf mass and luminosity (Yaron et al. 2005). 

\item
Recurrent novae have an extremely short supersoft X-ray emitting phase, compatible with very small 
envelope mass left. These objects - like RS Oph - are very challenging for theory, since they are 
only possible for a narrow parameter range (extremely large WD mass and quite large accretion rate). 
They are good scenarios for type Ia supernovae, since WD mass is expected to increase, but a caveat 
is that massive WDs are made of ONe and not of CO, and ONe WDs do not explode but rather collapse once 
they reach the Chandrasekhar mass (Guti\'errez et al. 1996).

\end{itemize}

\begin{figure}
\includegraphics[width=6cm,angle=-90]{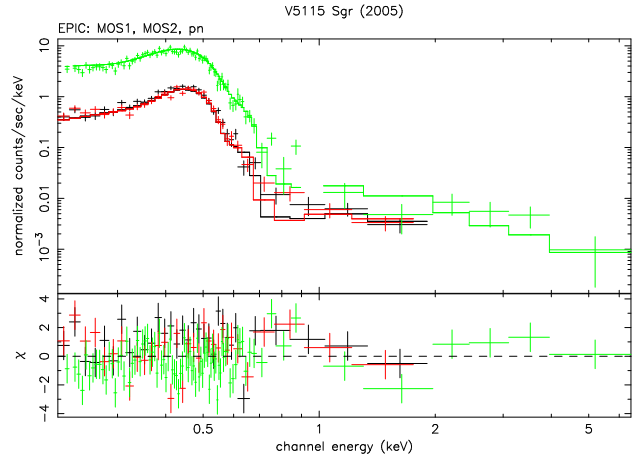}
\caption{Spectra of V5115 Sgr 2005a obtained with the EPIC (MOS1, MOS2 an pn) XMM-Newton 
cameras: normalised counts and residuals, for a ONe white dwarf atmosphere plus a thermal plasma 
best fit model. Model parameters: atmosphere has T$= 6\times 10^5$K and 
L$_{\rm WD}= 10^{36}$ erg/s, thermal plasma kT=2.5 keV and emission measure 
EM=$4 \times 10^{55}$ cm$^{-3}$, photoelectric absorption N$_{\rm H}=2 \times 10^{21}$ cm$^{-2}$. 
The adopted distance is 10 kpc and the total unabsorbed luminosity is L$\sim 10^{34}$ erg/s.}
\label{sgr05a}
\end{figure}

\begin{table}
\centering
\caption{Properties of some hydrodynamic models of nova explosions from 
Jos\'e \& Hernanz (1998) (see text for details).}
\label{tlab}
\begin{tabular}{lccc}\hline
Parameter                                   & ONe2 & ONe3 & ONe4\\
\hline
M$_{\rm wd}(M_\odot)$                       & 1.15   & 1.15    & 1.15\\
Mixing \%                                   & 25     & 50      &  75\\
X (H mass fraction)                         & 0.53   & 0.35    & 0.18\\
Z (metals mass fraction)                    & 0.27   & 0.50    & 0.74\\
$\Delta \rm M_{\rm env} (10^{-5}M_\odot)$   & 3.2    & 3.2     & 3.5\\
$\Delta \rm M_{\rm ejec} (10^{-5}M_\odot)$  & 2.3    & 1.9     & 2.6\\
$\Delta \rm M_{\rm rem} (10^{-5}M_\odot)$   & 0.9    & 1.3     & 0.9\\
L$_{\rm plateau} (10^4 L_\odot)$            & 4.3    & 4.9     & 5.4\\
$\tau_{\rm nuc}$ (yr)                       & 8.9    & 7.4     & 2.4\\
\hline
\end{tabular}
\end{table}

\acknowledgements

This research has been funded by the 
MICINN projects AYA 2008-01839/ESP, 
AYA 2008-04211-C02-01
and AYA 2007-66256, the AGAUR project 2009 SGR 315 
and by FEDER funds.


\end{document}